\documentclass[twocolumn]{aastex61}
\pdfoutput=1 
\usepackage{amsmath,amstext}
\usepackage[T1]{fontenc}
\usepackage{apjfonts} 
\usepackage[figure,figure*]{hypcap}
\usepackage{gensymb}
\usepackage{float}

\shorttitle{AASTeX 6.1 Template}
\shortauthors{A.Gibbs et al.}

\begin{document}

\title{VLT/SPHERE Multi-Wavelength High-Contrast Imaging of the HD 115600 Debris Disk: New Constraints on the Dust Geometry and the Presence of Young Giant Planets}

\author{Aidan Gibbs}
\affiliation{Steward Observatory, The University of Arizona, 933 N. Cherry
Avenue, Tucson, AZ 85721, USA}
\author{Kevin Wagner}
\affiliation{Steward Observatory, The University of Arizona,  933 N. Cherry
Avenue, Tucson, AZ 85721, USA}
\affiliation{NSF Graduate Research Fellow}
\affiliation{NASA Nexus For Exoplanetary System Science: Earths in Other Solar Systems Team}
\author{Daniel Apai}
\affiliation{Steward Observatory, The University of Arizona, 933 N. Cherry
Avenue, Tucson, AZ 85721, USA}
\affiliation{NASA Nexus For Exoplanetary System Science: Earths in Other Solar Systems Team}
\affiliation{Lunar and Planetary Laboratory, The University of Arizona, 1640 E. University Boulevard,
Tucson, AZ 85718, USA}
\author{Attila Mo\'or}
\affiliation{Konkoly Observatory, MTA CSFK, Konkoly-Thege Mikl\'os \'ut 15-17, H-1121 Budapest, Hungary}
\author{Thayne Currie}
\affiliation{Subaru Telescope, National Astronomical Observatory of Japan, 
650 North A`oh$\bar{o}$k$\bar{u}$ Place, Hilo, HI  96720, USA}
\affiliation{NASA-Ames Research Center, Moffett Field, CA, USA}
\affiliation{Eureka Scientific, 2452 Delmer Street Suite 100, Oakland, CA, USA}
\author{Mick\"ael Bonnefoy}
\affiliation{Max Planck Institute for Astronomy}
\author{Maud Langlois}
\affiliation{Centre de Recherche Astrophysique de Lyon (CRAL)}
\affiliation{Aix Marseille Universit\'{e}, CNRS, LAM (Laboratoire d 'Astrophysique de Marseille)}
\author{Carey Lisse}
\affiliation{Johns Hopkins University Applied Physics Laboratory}

\begin{abstract}

Young and dynamically active planetary systems can form disks of debris that are easier to image than the planets themselves. The morphology and evolution of these disks can help to infer the properties of the putative planets responsible for generating and shaping the debris structures. We present integral field spectroscopy and dual-band imaging from VLT/SPHERE ($1.0 - 1.7 \mu m$) of the debris disk around the young F2V/F3V star HD 115600. We aim to 1) characterize the geometry and composition of the debris ring, 2) search for thermal emission of young giant planets, and 3) in the absence of detected planets, to refine the inferred properties of plausible planets around HD 115600 to prepare future attempts to detect them. Using a different dust scattering model (ZODIPIC) than in the discovery paper (Henyey-Greenstein) to model the disk geometry, we find $a_0 = 46 \pm 2$ au for the disk's central radius and offsets $\Delta \alpha$, $\Delta \delta$ = $ -1.0 \pm 0.5, 0.5 \pm 0.5$ au. This offset is smaller than previously found, suggesting that unseen planets of lower masses could be sculpting the disk. Spectroscopy of the disk in Y-J bands with SPHERE shows reddish color, which becomes neutral or slightly blue in H band seen with GPI, broadly consistent with a mixed bulk disk composition of processed organics and water ice. While our observed field contains numerous background objects at wide separations, no exoplanet has been directly observed to a mass sensitivity limit of $2-3(5-7) M_{\rm J}$ between a projected separation of 40 and 200 au for hot (cold)-start models.
\end{abstract}


\section{Introduction}

Debris disks provide important clues to the late stages of planet formation (e.g., \citealt{Jang-Condell2015,Apai2015}). Second-generation dust disks can be produced by young, active planetary systems in which collisional cascades generate large amounts of dust, observable in infrared and longer wavelengths \citep{Wyatt2008,Gaspar2012}. Surveys have also shown that giant planets are more frequently found at large separations around stars with debris disks \citep{Meshkat2017}. 
Planets can interact with the disk to produce large-scale structures (rings, warps, spirals, etc.) that are often easier to observe than the planets themselves (e.g., \citealt{Su2013,Matthews2017,Currie2017}). 
Thus, stars with debris disks are not only good places to look for exoplanets, but disk structures can also be used to predict and constrain unseen planets for possible future detections and studies. 

The Scorpius-Centaurus OB association is an optimal target for disk studies, due to its proximity ($\sim$100-200 pc) and large collection of young systems. A high fraction of early-type Sco-Cen stars show strong infrared excess indicative of copious circumstellar dust \citep{Carpenter2009,Chen2014,Jang-Condell2015,Pecaut2016}. Sco-Cen includes many of the currently known directly-imaged planetary companions (e.g.\ \citealt{Rameau2013,Quanz2013,Bailey2014,Currie2015b,Chauvin2017,Keppler2018}) and numerous debris disks resolved in scattered light \citep[e.g][]{Thalmann2013,Currie2015,Lagrange2016,Matthews2017, Bonnefoy2017}.

HD 115600 (HIP 64995) is an F2V/F3V star located in the Lower Centaurus Crux region of the Sco-Cen OB Association whose debris disk could provide key insights into the planet formation. It has a measured distance of $109.6 \pm 0.5$ pc \citep{Brown2018} and is estimated to be $\sim$15 Myr old \citep{Pecaut2012, Pecaut2016}. The system was first observed to have an infrared excess by Spitzer/MIPS, with a fractional luminosity $L_{IR}/L_{*} = 1.7 \times 10^{-3}$ and dust mass 0.03 $M_{moon}$ \citep{Chen2011}, consistent with an extremely luminous dusty debris disk.
Using the Gemini Planet Imager, \cite{Currie2015} obtained the first images of the HD 115600 debris disk, resolving it into a bright ring-like structure. Given the system's age, the primary's mass and birth environment, and the disk's stellocentric distance, they hypothesized that the disk may be a good analogue for the early Kuiper belt. Its H-band spectrum is at least broadly consistent with predictions for a water ice composition, although other compositions could not be ruled out. By fitting an ellipse to the debris ring, they inferred a semi-major axis of $48 \pm 1.1$ au, an eccentricity of 0.1-0.2, and a disk offset of $\Delta \alpha, \Delta \delta = 0".018\pm 0".008, 0".029\pm 0".014$ ($2.0 \pm 0.9, 3.2 \pm 1.5$ au from the central star). To explore additional disk properties, \citet{Currie2015} created scattered light disk models using a Henyey-Greenstein phase function and forward-modeled them through their reduction pipeline, finding neutral scattering was the best fit. Furthermore, they found that the debris disk is very thick, with a width-to-mean radius ratio of $\Delta r/r_0 \sim 0.37$. 

The sharp ring-like structure of the disk may suggest sculpting by a planet (or planets) and the measured offset of the disk further implies that this planet could be massive. Numerical modeling by \cite{Thilliez2017} determined that disk structures driven by a  $\sim 8  M_J$ planet located at $\sim$30 au are the best fit to observed morphology. While the predicted contrast ($4 \times 10^{-5}$ in H band) of this hypothetical planet is within the theoretical performance capabilities of current extreme adaptive optics (hereafter extreme AO) instruments such as SPHERE and GPI \citep{Mesa2015,Bailey2016}, the orientation of the disk (i = $79^\circ$) makes the detection very challenging.  

By analyzing new multi-wavelength extreme AO observations of HD 115600's debris disk with updated models, we can better constrain the disk's geometry, provide new constrains on the system's inventory of massive planets possibly sculpting the disk, and provide new insights into the disk's composition. Numerous debris disk studies including \citet{Currie2015} interpret scattered light images using the standard, simple Henyey-Greenstein (H-G) scattering phase function.  In addition to noting that H-G functions are ad hoc, recent results further challenge their accuracy \citep{Milli2017,Hughes2018,Goebel2018}. Empirically derived scattering functions of debris disks more closely resemble a combination of Henyey-Greenstein functions, similar to the phase function of zodiacal light \citep{Graham2007,Ahmic2007}. Disk modeling with these functions, rather than H-G, may lead to a revision in disk properties. Additionally, spectroscopy covering a wider wavelength range may clarify whether HD 115600's disk composition shows evidence for a single constituent or has mixed dust composition. 


In this study, we present new, multi-wavelength images and spectra of the debris disk around HD 115600 taken with VLT/SPHERE. The processed data are presented and described in Section \ref{sec:reduction}. These images provide a more detailed look at the disk structure as well as expanded wavelength coverage in the $Y-$ and $J$-bands, and the results are discussed in Section \ref{sec:results}. To better constrain the disk structure, we model the disk as an optically thin scattering ring using a scattering function derived from zodiacal light \citep{Hong1985}, and present our model in Section \ref{sec:analysis}, with a discussion in Section \ref{sec:discuss}. In Section \ref{sec:spectra}, we examine the new YJ spectra  in combination with the H-band data from \citet{Currie2015}. Finally, we provide new limits on the mass and orbit of any unseen exoplanets in Section \ref{sec:planets}, and conclude with a summary of our findings in Section \ref{sec:sum}. 

\section{Observations and Data Reduction}
\label{sec:reduction}

HD 115600 was observed on 2015 June 04 as part of ESO Program 095.C-0298, using the Spectro-Polarimetric High-contrast Exoplanet Research (VLT-SPHERE: \citealt{Beuzit2008}) instrument. The observations were conducted with the IRDIFS mode which allows operation of the infrared dual-band imager and spectrograph (IRDIS: \citealt{Dohlen2008,Vigan2010}) with the H2 and H3 filters (central wavelengths $1.576\pm 0.052$ and $1.667\pm 0.054$ $\mu m$ respectively), simultaneously  with the integral field spectrograph (IFS: \citealt{Claudi2008,Mesa2015}) in YJ bands ( $0.95-1.35 \mu m$) with a spectral resolution of $R \sim 50$. SPHERE IFS has a narrow field-of-view of $\sim 1$ arcsecond in radius, with a plate scale of 7.46 mas/pixel, roughly 0.8 au per pixel at the distance of HD115600. IRDIS has a wide field-of-view of $11" \times 12.5"$, with a lower resolution plate scale of 12.25 mas/pixel \citep{Claudi2008, Dohlen2008,Vigan2010}. 

During the observations, the SAXO extreme adaptive optics system \citep{Fusco2006} corrected for wavefront errors, and the apodized Lyot coronograph, which has an occulting mask with radius of 0$\farcs$0925, attenuated the stellar light contribution. Before the observations, a short star-centering sequence was obtained by applying a sinusoidal pattern to the deformable mirror \citep{Langlois2013}. This deformation creates 4 satellite spots in a diagonal cross pattern around the star that can be used to precisely locate the star center behind the coronagraph. Using this method, the uncertainty in the star center is 0.25 pixels in both the x and y directions \citep{Mesa2015}. Flux calibrations were obtained before the observation by slewing the star off of the coronagraph while inserting a neutral density filter to avoid saturation. SPHERE obtained 64 images for IFS and and 64 for IRDIS with exposure times of 64 seconds each, for a total integrated exposure time of 68 minutes for each of the two instruments. The instrument field derotator was switched off for the observations to allow angular differential imaging in post-processing, and the field rotation during the observation sequence was $\sim 27^{\circ}$. The key parameters of the observation are summarized in Table 1. 

The basic reduction steps for IFS data, including dark current subtraction, flat field division, microspectra extraction, and wavelength calibration, were performed using a combination of the ESO SPHERE pipeline (version 0.15.0, \citealt{Pavlov2008}) and the custom IDL tools described in \citealt{Vigan2015}. These tools add some additional pre-processing steps beyond those of the pipeline, such as bad pixel correction using the IDL procedure MASKINTERP, and cross-talk correction (spurious light contamination between integral field unit lenslets). Small scale cross-talk is corrected using a $41\times 41$ kernel with a Moffatt function to reduce the presence of doubles to the aforementioned satellite spots, which are a result of cross-talk. To reduce the stellar halo and speckle noise which remains after extreme AO coronagraphy, we utilized both angular differential imaging (ADI; \citealt{Marois2005}) and spectral differential imaging (SDI; \citealt{Racine1999}) techniques. These methods utilize varying field rotation and wavelength respectively to separate astrophysical objects from the speckle noise. We used our own principal component analysis (PCA) based Karhunen Lo\'eve image projection (KLIP; \citealt{Soummer2012}) algorithms of ADI and SDI, which are described in \citet{Hanson2015} and \citet{Apai2016}. We used annular PCA, with 15 annuli, 14 pixels in width each, and six angular segments. We retained 7 KL basis vectors for ADI and SDI, with a minimum field rotation of $0.5 \times FWHM$ for ADI, and a minimum speckle radial movement of $1.5 \times FWHM$ for SDI between images. Before the PSF subtraction, we applied a sharpening high-pass filter with a width of 11 pixels to the data to reduce low-frequency noise. The reduced images were stored in x-y-$\lambda$ (Cartesian and wavelength) cubes, as well as band-averaged flux density images in the Y- and J- bands, which greatly improves signal to noise (S/N) compared to the detection in the individual IFS channels. The first seven frames of the cube were not included in the band-averaged images to eliminate frames impacted by detector persistence from the flux calibration frames. All images for a respective wavelength were combined with a noise-weighted mean \citep{Bottom2017}. Basic reduction steps for IRDIS data were similarly performed using custom IDL tools for bad pixel correction, dark current subtraction, flat fielding, and star centering. We again used classical and KLIP ADI and SDI techniques for PSF subtraction.  

\begin{deluxetable*}{ccccccccc}[t]

\tablecaption{HD 115600 SPHERE Observations}

\tablehead{\colhead{Date} & \colhead{Instrument} & \colhead{Filter} & \colhead{Total Exposure Time} & \colhead{Sub-Int. Time} & \colhead{\#Int.} & \colhead{Average Seeing} & \colhead{Average Airmass} & \colhead{} } 

\startdata
June 4 2015 &  IFS &   YJ (0.96 - 1.34 $\mu m$) &   $\sim 68$ min &    64.0 s & 64 &   1.1" &   1.22 \\
June 4 2015 &   IRDIS &   H23 (1.593 - 1.667 $\mu m$) &   $\sim 68$ min &   64.0 s & 64 &  1.1" &   1.22 \\
\enddata



\label{tab:obs}
\end{deluxetable*}

\section{Results}
\label{sec:results}

\begin{figure*}[p]
\centering
\includegraphics[width=0.6\textwidth]{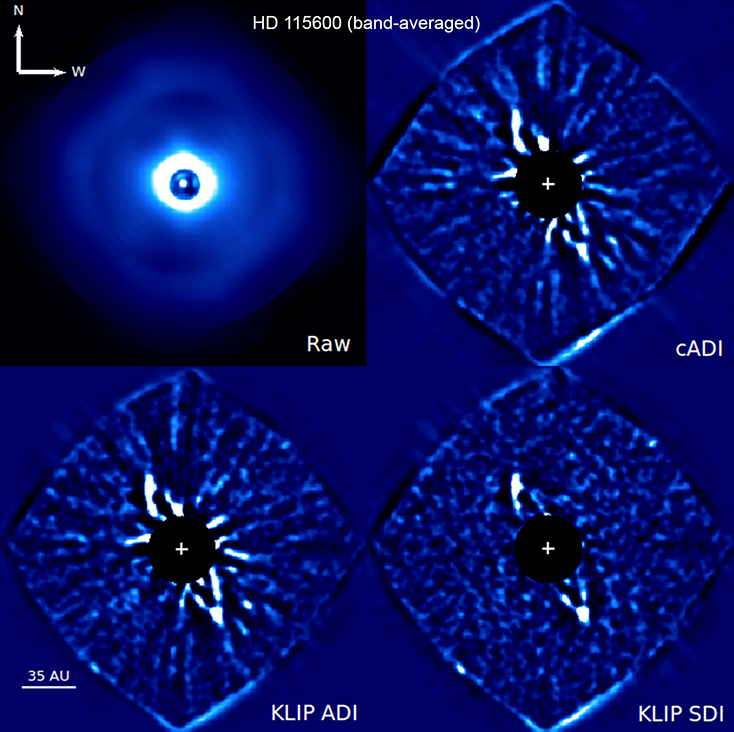}
\caption{\textbf{Band-averaged (0.96 - 1.34 $\mu m$), IFS images of the HD 115600 disk.} The raw image is presented in the top left. ADI and SDI reductions are presented in the other 3 panels with high-pass filtering. The disk is clearly visible with ADI and SDI techniques. The image center is marked by a cross in each image, and noise from the central star is masked to $\sim 25$ au.}
\label{fig:ifs}
\end{figure*}

\begin{figure*}[p]
\centering
\includegraphics[width=0.9\textwidth]{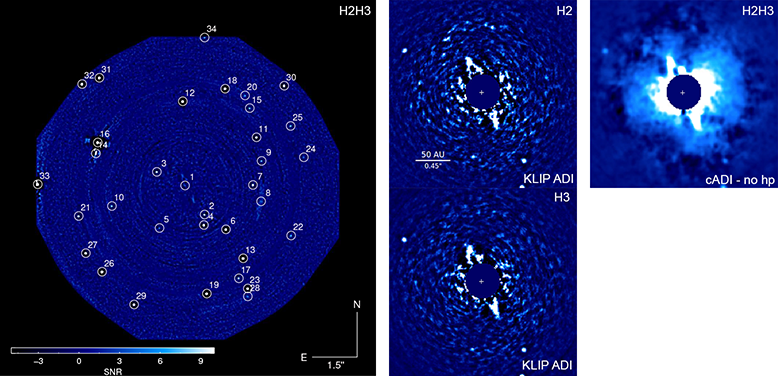}
\caption{\textbf{IRDIS dual-band (H23, 1.587 and 1.667 $\mu m$) images of the HD 115600 disk.} Left: Signal-to-nose map of the full field-of-view (11 arcseconds) with detected point-sources labeled. The point labeled 1 is a disk feature. Center: KLIP ADI reductions with high-pass filtering for H2 (top) and H3 (bottom) in a region of 2.5" around the disk. Right: Classical ADI reduction with no high-pass filtering for the same region in both wavelengths. Noise from the central star is masked in inset images to $\sim 25$ au with the star center marked by a cross. Note that the brightness scaling changes between left, center, and right.}
\label{fig:irdis}
\end{figure*}

We detect the disk in the band-integrated flux density IFS images, shown in Figure \ref{fig:ifs}, with a radial extent from r$\sim$0."35 to 0."48, corresponding to a projected distance of $\sim$ 39 to 53 au. The disk is also visible in IRDIS images, presented in Figure \ref{fig:irdis}. To estimate the signal-to-noise ratio of the disk, 
we calculate the noise as the standard deviation of the band-integrated image in circular apertures with widths corresponding to the diffraction limited, full-width half maximum (FWHM, $\sim 9$ pixels), which do not overlap with the regions where the disk is visible, but are centered at similar radii to disk ansae ($\sim$55 pixels from image center). The signal is defined as the average pixel value of two circular regions of FWHM radius centered on the north and south disk ansae where signal is highest (center pixels 115x, 190y and 165x, 90y). By this metric, with SPHERE-IFS we achieve a peak SNR$\sim$20 in the disk ansae, decreasing in disk regions with smaller separations.   \citet{Currie2015} report a peak SNR of $\sim$ 8 in GPI data, where they do not mask the disk or self-subtraction footprints in determining the noise.   Using the same approach for calculating noise, the disk's SNR in the SPHERE data is slightly higher ($\sim$ 10\%) than in GPI data. 


To look for any additional structure that may be present outside of the disk, we also performed PSF subtractions without high-pass filtering, which can suppress extended and faint features. We did not find evidence for any new disk features. Within a projected distance of $\sim 25$ au from the star, which is masked in Figures \ref{fig:ifs} and \ref{fig:irdis}, residual speckle noise dominates the image and we are not able to probe disk structure reliably.

Multi-wavelength coverage is used to discriminate between speckle noise, which scales with wavelength, and point-source detections, which do not. We did not detect any point-sources in the vicinity of the disk within 1" of the star. Point-sources detected in the IRDIS field radially closer than the point labeled 4, shown in Figure \ref{fig:irdis}, exhibit motion visually consistent with background stars due to the proper motion of HD 115600 between GPI and SPHERE image epochs. Multiple epoch data was not attainable for point-sources radially further out than point 4. Appendix A includes the properties for all detected sources in the SPHERE and GPI fields. In Section \ref{sec:planets}, we will quantify and discuss the contrast and mass-sensitivity limits these observations place on the potential presence of planets in the system based on hot and cold start core-accretion models of planet formation and evolution.

\section{Disk Models}
\label{sec:analysis}

Our aim with disk modeling is to identify the simplest physically viable model of the disk, accounting for projection and light-scattering effects. By understanding the fundamental geometry, we can later attempt to make predictions about disk sculpting by possible exoplanets. We perform the disk modeling by two independent methods, ellipse fitting and with the model ZODIPIC, in order to asses the accuracy of both. We make no attempt to model the disk at the wavelengths of IRDIS, given the benefits of the IFS data such as extended wavelength coverage compared to previous studies and the smaller inner working angle.

We performed primary modeling of the data using the model ZODIPIC \citep{Kuchner2012}, originally written to model exozodiacal dust, but which is generally applicable to optically thin debris disks. ZODIPIC is based on the radiative model of zodiacal dust in \cite{Kelsall1998}, derived from all-sky data collected by the Cosmic Background Experiment/Diffuse Infrared Background Experiment (COBE/DIRBE). This model was chosen not only for ease-of-use, but because the scattering phase function for zodiacal light \citep{Hong1985} contained in ZODIPIC resembles those empirically derived from debris disks \citep{Graham2007,Ahmic2007}. This setup allows us to control the disk size, inclination, offset, and scattering phase function, among other parameters. More complex radiative transfer models could be used, but were not warranted by the data quality and the goals of our study.

The general procedure we used to find a best-fit disk model with ZODIPIC was to produce an evenly spaced grid of disk models with varying geometric parameters, and to subtract them from the raw data to find the model that minimizes the residual disk signal in our processed images without considering noise. The residual was calculated for every model in the grid to avoid local minimums and find the absolute lowest value. ZODIPIC parameters related to the star and instrument performance were kept constant through all models. These include the stellar type, distance ($\sim 109.6 \pm 0.5 pc$), and pixel-scale for SPHERE IFS. ZODIPIC also required the observation wavelength as a parameter, therefore we produced a separate model in each IFS wavelength channel for every geometry. The phase function parameter we used is the standard function included in ZODIPIC from \citet{Hong1985} for wavelengths below 4.2 microns, which covered scattering between 30 and 180 degrees. We do not alter or explore the scattering phase function in the interest of keeping a manageable grid size. The geometric parameters varied within our grid search were position angle, inclination, disk radius, and projected location of the disk's center. We set the inner and outer disk radius to the approximate visual size, 39 to 53 au, and did not vary these parameters, first to reduce the grid dimensions, and also because we found their variance produced negligible effects in the subtraction residuals. The grid was centered on the best-fit parameters determined by \citet{Currie2015} using ellipse fitting, which we also performed as a secondary analysis. Table 2 contains our range and step size for the varied parameters. 

To subtract a geometric model from our data, we first convolved the model with the measured SPHERE PSF. The width of this PSF was measured from the flux calibration data cube in each wavelength channel. The convolved model was then multiplied by a brightness scale factor, which was initially the same for every wavelength, to bring the brightness of the model to the same approximate brightness as the disk. We subtracted the convolved model from our raw data, and processed it with classical ADI. Classical ADI was chosen to cover a larger parameter space since PCA-based KLIP routines are more computationally expensive. After processing, the disk residuals were measured in the band-averaged images in two circular apertures of 40 pixel ($\sim 33$ au) diameter centered on each disk ansa. These apertures were chosen to avoid strong speckle noise near the image center, but to be large enough to accommodate the range of disk sizes and angles in our grid. The geometry that produced the least-squared residual was adopted as our best-fit. The error is estimated as twice the step size, that is twice the smallest difference between parameter values in our grid, for all grid parameters. The uncertainty in the star centering ($0.2$ au in x and y) is also considered, but is a minor contribution compared to the step size.

Our final step was to find the optimal brightness of the geometric model in each wavelength channel to further minimize the disk residual after subtraction. We subtracted the geometric model in each channel at varying brightnesses similarly to before, but measured the disk residual in the corresponding channel of the processed data cube instead of the band-averaged image. The brightness that minimizes the residual in a channel is adopted as the best-fit for that channel. The error again was estimated as twice the final step size, the final difference between neighboring brightness values we tried, which is 1/16 of the best-fit brightness value. 

\begin{figure*}
\centering
\includegraphics[width=0.75\textwidth]{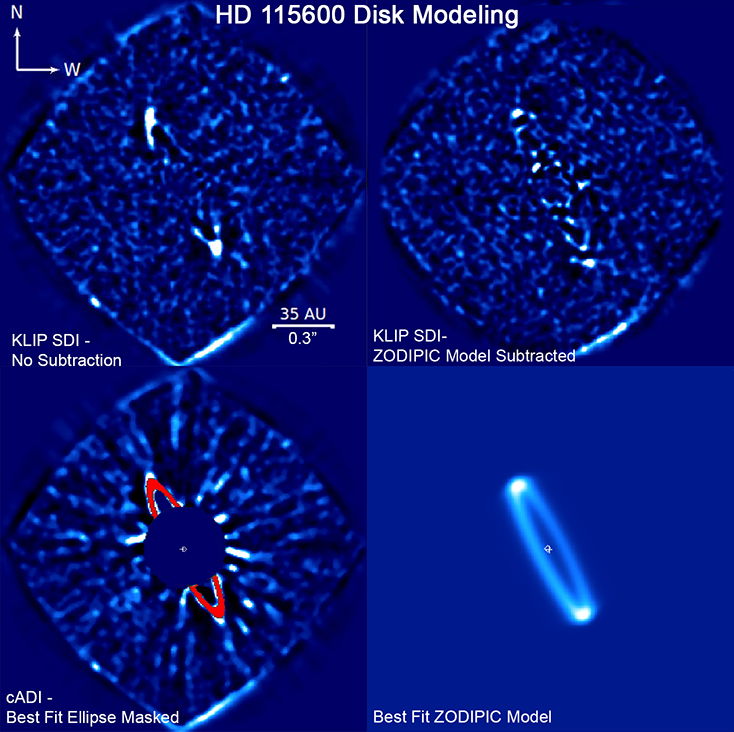}
\caption{\textbf{Disk Modeling.} Top Left: Band-averaged, KLIP SDI reduction of IFS images without subtraction. Top Right: The same reduction with the best-fit disk model from ZODIPIC subtracted. Bottom Left: Best fit ellipse model in green over classical ADI reduction. Noise from the central star is masked to $\sim 25$ au. Bottom Right: ZODIPIC best-fit model which is used for subtraction. The star center marked by a cross. The centers of the respective disk models are marked with diamonds.}
\label{fig:subtraction}
\end{figure*}

Figure \ref{fig:subtraction} shows a KLIP SDI reduction of the disk in IFS, the best-fit model disk produced by ZODIPIC, and the KLIP SDI reduction of the raw data with the best-fit model subtracted. KLIP SDI was chosen to display the before and after disk model subtraction as it has the least speckle noise, even though SDI processing was not used for the actual disk modeling. The best-fit disk model was only processed with SDI for display. With ZODIPIC, we found the best-fit model has a position angle and inclination of $P.A.=27.0\degree \pm 1.0$ and $i=80.0\degree \pm 1.0$. Our model estimated the disk radius at $46.0 \pm 2.0$ au and that the projected center of the disk is minimally offset from the central star, with best-fit $\Delta x, \Delta y = -1.0 \pm 0.5, 0.5 \pm 0.5$ au.

Since we did not directly probe the inner and outer disk radius with our ZODIPIC model, we must consider what the disk radius indicates further. Scattered light observations mostly probe those small grains that can  
strongly be affected by interaction with stellar radiation forces or wind. 
Pushed onto eccentric or unbound orbits by the radiation pressure these 
small particles can form an extended halo outside the ring of parent planetesimals.
Such halos can make difficult the accurate determination of the planetesimal 
belts' outer edge in scattered light observations, however we can reasonably assume 
that the peak of the surface brightness distribution is coincided with the planetesimal belt 
thereby probing its location.
Though the presence of gas can alter this picture in special cases \citep{richert2018}, sensitive observation 
of HD\,115600 with the ALMA interferometer found no gas \citep{Lieman-Sifry2016}, therefore in the following the belt is assumed to 
be located at a radius of 46\,au. 

We performed a secondary, independent geometric analysis using the same ellipse fitting method used originally by \citealt{Currie2015} on HD 115600 and described in \citealt{Thalmann2011}. First, we applied a median filter (FWHM, $\sim 4$ pixels width) to the classical ADI image of the disk to reduce the pixel-to-pixel noise. We then constructed a large grid of ellipses ($\sim 10,000$) around the best-fit disk model as determined by ZODIPIC and find a new best-fit geometry as the ellipse which contains the highest average signal within its trace. The center of the image ($\sim 25 $ au) was again masked out for this analysis to avoid contributions from the strong speckle noise.

The best-fit ellipse trace is shown superimposed over the disk in Figure \ref{fig:subtraction}. It has projected semi-major and minor axes of $a = 44.5 \pm 0.8, b = 8.9 \pm 0.8$ au corresponding to an inclination of $i=78.5^{\circ} \pm 1.0$, with a position angle $P.A. = 27.5^{\circ} \pm 1.1$. The disk center is offset $\Delta x, \Delta y = 0.8 \pm 0.6, 0.0 \pm 0.6$ au. Uncertainty is again estimated as twice the step size. We also perform the same analysis on KLIP ADI and SDI images. In those cases, we achieve consistent results, and the estimated disk offset is actually smaller.


\begin{deluxetable*}{ccccc}




\tablecaption{ZODIPIC Disk Modeling}


\tablehead{\colhead{Parameter} & \colhead{Range Tested} & \colhead{Step Size} & \colhead{Best Fit} & \colhead{Uncertainty} }

\startdata
Disk Radius (au) & 45 to 49 & 1.0 & 46 & $\pm 2.0$ \\ 
    Position Angle (deg) & 23 to 29 & 0.5 & 27 &$\pm 1.0$ \\
    Inclination (deg) & 78 to 81 & 0.5 & 80 & $\pm 1.0$ \\
    x Offset (au) & -1 to 4 & 0.25 & -1.0 & $\pm 0.5$ \\ 
    y Offset (au) & -1 to 4 & 0.25 & 0.5 & $\pm 0.2$ \\ 
\enddata



\label{tab:zoditab}
\end{deluxetable*}

\section{Discussion}
\label{sec:discuss}

\subsection{Disk Offset}

A disk pericenter offset can be caused by gravitational perturbations from one or more exoplanets in close proximity to the disk, and thus is an indication that unseen planets may be present. Many other disks in Sco-Cen have shown structures and asymmetries, including HD 106906 \citep{Kalas2015}, HD 111520 \citep{Draper2016}, HD 110058 \citep{Kasper2015}, HIP 67497 \citep{Bonnefoy2017}, etc.

Both disk modeling methods individually indicate there may be a small ($\sim 1$ au) projected disk offset in the x-direction, but do not agree on which direction, and are consistent with zero offset within $2\sigma$ errors. Offsets on the order of $\sim 1$ au are at the scale of one IFS pixel, and are therefore difficult to determine precisely. Thus, we are only able to constrain that the projected offset does not exceed $\sim 1.5$ au in either direction.

Our results on the disk eccentricity imply that any unseen exoplanets are likely less massive than previously estimated. Using the projected disk offset estimated by \cite{Currie2015}, \cite{Thilliez2017} showed that a 7.8 $M_J$ planet could be present in the system at 30 au with an eccentricity of $e=0.2$, which could be detected by SPHERE or GPI. Due to the inclination of the system, however, there is a high probability that the planet would lie within the inner working angle of both instruments, and thus undetected.  If the disk eccentricity is smaller, though, planets far less massive than $7.8 M_J$ may be able to sculpt the disk, plausibly low enough to elude detection in our data and those in \citealt{Currie2015}.


\subsection{Stirring of the Disk}

Planetesimals must be stirred to produce smaller and smaller fragments and finally dust via their 
collisions \citep{matthews2014}. This stirring could either be related to large planetesimals embedded in the belt 
\citep[{self-stirring,}][]{kb2008} or a giant planet(s) located somewhere in the system 
\citep[{planetary stirring,}][]{mustill2009}. In the following we will explore whether 
the self-stirring scenario is feasible in the HD\,115600 system or we need a giant planet 
to explain the dust production. In the classical self-stirring model collisional coagulation among smaller bodies leads to the formation of Pluto-sized ( r $\sim$1000\,km) 
planetesimals that then can ignite a collisional cascade in the belt via their dynamical perturbation \citep{kb2008}. The larger the radial distance and the smaller the surface density of the disk, the longer is the time needed for the buildup of these large bodies. \citet{kb2008} provided an analytical formula for this 
timescale: 
$t_{\rm 1000} = 145 x_{\rm m}^{-1.15} (a/80\,{\rm au})^3 (2M_{\sun}/M_*)^{3/2} [\rm Myr]$. 
Here $a$ is the radial distance, while the $x_{\rm m}$ parameter scales with the initial 
surface density, in the case of the minimum-mass solar nebula $x_{\rm m} = 1$. Adopting the 15 Myr estimated age of HD115600 as the available time for the formation of Pluto-sized bodies and a primary stellar mass of 1.5\,M$_\odot$ \citep[taken from][]{Chen2014}, 
we found that a disk with $x_{\rm m}\sim2.5$ is enough to explain the stirring at 46\,au.  
Alternatively, according to turbulent concentration and gravitational clumping
models \citep{johansen2007,cuzzi2008} large planetesimals can form directly from the 
concentration of small pebbles in a protoplanetary disk. 
Though these models predict very rapid formation of hundred kilometer size bodies even at 
radii $>$100\,au \citep{carrera2017}, additional time is needed to excite the neighbouring 
disk sufficiently \citep{krivov2018}. According to equation~34 from \citet{krivov2018} 
in this model the stirring of the belt at a time of 15\,Myr needs an even lower initial 
surface density ($x_{\rm m}\sim0.3$) than in the case of the gradual buildup approach.
Obviously, the initial surface density could not be arbitrarily large: according to 
\citet{mustill2009}, $x_{\rm m} > 10$ would imply that the self-stirring scenario may not be feasible in the given system raising the suspicion that the dynamical excitation is rather related to a planet.  
The obtained low $x_m$ values thus indicate that self-stirring can work 
 in the HD\,115600 systems and it is not necessary to assume the 
presence of a giant planet. 


\subsection{Analysis of the Spectral Energy Distribution}

The spectral energy distribution (SED) of a stellar system can be used to determine if there are one or more circumstellar disk components based on the number of temperature peaks. If a disk has two spatially distinct components, the inner component will be warmer than the outer component and both will have unique blackbody distributions visible in the SED. Examining the SED of HD 115600, we can determine if there may be an inner disk component not visible in our data.

By modeling the SED of HD~115600, \citet{Chen2014} 
proposed a two-temperature disk model with $T_{\rm d,warm} = 499$\,K and 
 $T_{\rm d,warm} = 109$\,K to fit the observed mid- and far-infrared (far-IR)
 excess. Similar analyses of \citet{Ballering2013} and more recently 
 \citet{Currie2015}, however, led to single component models at temperatures of 109 and 115\,K, respectively. All of these results were based on measurements shortward of 70 {\micron}. In order to better characterize the far-IR SED of our target we examined unpublished 100 and 160 {\micron} observations obtained with the PACS instrument onboard the Herschel Space Observatory. 
 We used the Herschel Interactive Processing Environment 
 \citep[HIPE v13,][]{Ott2010} to process the data 
 following the basic steps described in \citet{Moor2015}.
HD\,115600 is detected at both wavelengths, however at 160{\micron} 
it is contaminated by surrounding extended emission 
therefore no photometry was derived. At 100{\micron} our aperture photometry 
yielded a flux density of 138$\pm$12\,mJy. The aperture radius was set to 
6{\arcsec}, the sky background was measured in an annulus between 40{\arcsec} and 50 
{\arcsec}. The quoted uncertainty is a quadratic sum of the measurement 
and calibration \citep[7\%,][]{Balog2014} errors. 
To compile the final SED, previously obtained infrared photometric and spectroscopic data taken from the literature and public databases 
\citep{wright2010,ishihara2010,lebouteiller2011,Chen2014} were supplemented by 
our new 100{\micron} and a recent 1.24\,mm \citep{lieman-sifry2016} photometry 
of the source. To estimate the photospheric contribution at relevant 
wavelengths an ATLAS9 model \citep{castelli2003} was fitted to the
optical and near-IR photometric data of HD\,115600 
\citep[taken from][]{perryman1997,hog2000,cutri2003}. 
Based on our analysis of the source's Str\"omgren photometry \citep{paunzen2015} 
the influence of reddening was neglected, while for the metallicity
 a value of [Fe/H] = $-$0.16 was derived using the method proposed by \citet{casagrande2011}.
By fixing [Fe/H] = $-$0.16 as well as {\ensuremath{\log g}} = 4.25 (i.e. assuming the primary is a young dwarf star of slightly subsolar metallicity) a good best fit photospheric model was achieved using {\ensuremath{T_{\mathrm{eff}}} = 6850\,K, consistent with an F2V star (Mamajek \url{http://www.pas.rochester.edu/~emamajek/EEM_dwarf_UBVIJHK_colors_Teff.txt}).
Figure~\ref{fig:sed} displays the SED and the obtained photosphere model of our target. 
       
After the subtraction of the photospheric contribution, the excess SED was fitted 
by 1) a single temperature modified blackbody model where the emissivity is 1 at
wavelengths shorter than 100{\micron} and varies as $(\lambda / 100{\micron})^{\beta}$ 
at longer wavelengths, and 2) by a two-temperature model where the single modified 
blackbody component was complemented with a warmer simple blackbody. 
We utilized a Levenberg-Marquardt algorithm \citep{markwardt2009} to find the best-fitting model.
In good accordance with \citet{Ballering2013} and \citet{Currie2015} we found 
that the SED is adequately described by a single modified blackbody with 
 $T_{\rm dust} = 111\pm3$\,K, $\beta = 0.92\pm0.22$, and $L_{\rm disk}/L_{\rm bol} = 0.002$.

\begin{figure*}[t]
\begin{center}
\includegraphics[width=0.75\textwidth]{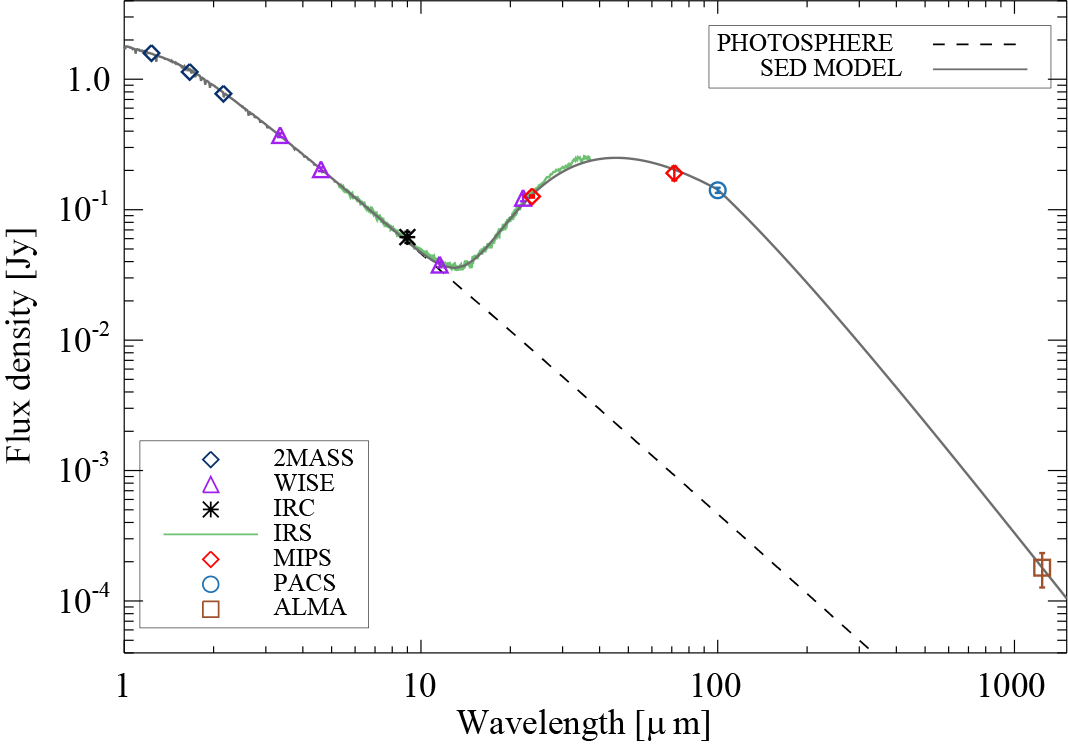}
\end{center}
\caption{Spectral energy distribution of HD\,115600 overplotted by the
fitted modified blackbody model. 
\label{fig:sed}}
\end{figure*}

\subsection{Dust-rich Debris Disks around F-G-type Sco-Cen members}

HD\,115600 is a member of the Sco-Cen association. 
To put into context our results on its disk parameters in the following we 
make a comparison with other dust-rich debris disks in the same group.
 Using {\sl Spitzer} 24$\mu$m photometric data \citet{Chen2011} inferred an excess 
fraction of $\sim$33\% for F-G-type members of the Sco-Cen association. 
Eleven out of the revealed disks were found to exhibit fractional luminosity higher than $10^{−3}$, and ten from this bright disk subsample, including HD 115600, harbored dust colder than 150 K. 
Observations of these ten dust-rich systems with ground-based high-contrast imaging 
systems such as SPHERE and GPI in scattered light and with the ALMA interferometer at millimeter wavelengths resulted in spatially resolved disk images in all but one case
 (HD\,117214).
Fundamental properties of these objects, derived based on their resolved images and 
analysis of their SED, are summarized in Table~\ref{tab:scocendebris}.

\begin{deluxetable*}{lcccccccccc}
\tablecaption{Stellar and Disk Parameters \label{tab:scocendebris}}
\tablecolumns{11}
\tabletypesize{\scriptsize}
\tablehead{
\colhead{Name} & \colhead{Spt.} & \colhead{Dist.} & \colhead{Lum.} & \colhead{Group} & 
\colhead{$L_{\rm disk}/L_{\rm bol}$} & \colhead{$T_{\rm dust}$} & \colhead{Ref.} &
\colhead{$R_{\rm BB}$} & 
\colhead{$R_{\rm belt}$} & \colhead{Ref.}\\
\colhead{} & \colhead{} & \colhead{(pc)} & \colhead{(L$_\odot$)} & \colhead{} & 
\colhead{$\times$10$^{-3}$} & \colhead{(K)} & \colhead{} & \colhead{(au)} & 
\colhead{(au)} & \colhead{}
}
\colnumbers
\startdata
HD\,106906  & F5V     & 103.3 & 6.56 & LCC & 1.8 &  53$^{*}$ & 1 & 70.5 & 71 & 4 \\
HD\,111520  & F5/F6V  & 108.9 & 2.58 & LCC & 1.5 &  49$^{*}$ & 3 & 51.7 & 70 & 2 \\
HD\,114082  & F3V     &  95.7 & 3.86 & LCC & 4.3 & 110 & 1 & 12.5 & 31       & 7 \\ 
\bf{HD\,115600}  & \bf{F2IV/V}  & \bf{109.6} & \bf{4.84} & \bf{LCC} & \bf{2.0} & \bf{111} & \bf{4} & \bf{13.8} & \bf{46}       & \bf{8} \\
HD\,120326  & F0V     & 113.9 & 4.51 & UCL & 2.2 & 114 & 1 & 12.6 & 62       & 1 \\
HD\,129590  & G3V     & 136.0 & 2.98 & UCL & 7.7 &  89 & 1 & 16.9 & 57       & 6 \\
HD\,145560  & F5V     & 120.4 & 3.24 & UCL & 3.9 &  53$^{*}$ & 1 & 49.5 & 79$^{*}$ & 5 \\
HD\,146181  & F6V      & 125.0 & 2.73 & UCL & 3.0 &  71$^{*}$ & 1 & 25.3 & 79$^{*}$ & 5 \\
HD\,146897  & F2/F3V  & 131.5 & 3.29 & US  & 5.6 &  94$^{*}$ & 2 & 15.9 & 78 & 3 \\
\enddata
\tablecomments{
Column (1): Target name. 
Column (2): Spectral type (from SIMBAD). 
Column (3): Distance \citep[from {\sl Gaia DR2} parallax,][]{lindegren2018}.
Column (4): Luminosity. Data were taken from {\sl Gaia DR2} catalogue \citep{lindegren2018,andrae2018}
except for HD\,115600 where we used our own estimate. 
Column (5): Group membership. LCC: Lower Centaurus Crux association; UCL: Upper Centaurus
	Lupus association; US: Upper Scorpius association.
Column (6): Fractional luminosity of the disk.
Column (7): Dust temperature. For two-temperature disks (marked by asterisks) 
the temperature of the colder component was quoted.
Column (8): Reference for dust temperature. (1) \citet{Ballering2013}; (2) \citet{Chen2014}; 
(3) \citet{Draper2016}; (4) this work.
Column (9): Disk radius estimated from the characteristic dust temperature 
assuming grains act like blackbodies ($\frac{R_{\rm BB}}{au} = \left(\frac{L_{\rm star}}{L_{\odot}}\right)^{0.5} \left(\frac{278\,K}{T_{\rm disk}}\right)^2$ ).
Column (10): Disk radius derived from spatially resolved scattered light or millimeter 
images. For scattered light data 
the radius of peak dust density was adopted, in the case of millimeter 
data (marked by asterisks) the average between the inner and outer belt radii were used. 
The original estimates (for references see Col. 11) were recomputed considering the 
new {\sl Gaia DR2} distances.
Column (11): Reference for $R_{\rm belt}$ values: (1) \citet{Bonnefoy2017}; (2) \citet{Draper2016}; 
(3) \citet{engler2017}; (4) \citet{Lagrange2016}; (5) \citet{Lieman-Sifry2016}; 
(6) \citet{Matthews2017}; (7) \citet{wahhaj2016}; (8) this work.
}
\end{deluxetable*}

The radii of the resolved dust belts range between 30 and 80\,au. With its radius of 46 au, HD 115600 is the second smallest object after the disk of HD 114082 and resembles our Kuiper-belt the most among all the targets \citep[see also][]{Currie2015}. 
The ratio of the disk sizes derived from the spatially resolved images ($R_{\rm belt}$) to radii
inferred from dust temperatures assuming large grains behaving as blackbodies ($R_{\rm BB}$) is higher 
than 1 for all cases. This is a well known signature of the presence of small dust particles that, as being inefficient emitters at long wavelengths, are hotter than blackbodies \citep{Booth2013}. Small grains should not be stable under the influence of radiation pressure, and their presence suggests an active source of supply from grinding/aggregating planetesimals and sublimating icy bodies.

Figure~\ref{fig:gamma} shows the $\Gamma=R_{belt}/R_{BB}$ ratios as a function of stellar luminosity 
for our targets.  For comparison, debris disks resolved with the Herschel Space Observatory around stars having luminosity similar to that of our study are also displayed \citep[data for these objects were taken from table~4 of][]{pawellek2014}.  
As Figure~\ref{fig:gamma} demonstrates dust-rich debris disks around F-G-type members of Sco-Cen show 
a great variety in terms of 
$\Gamma$ values. On the one hand HD\,106906 with its $\Gamma$ value close to 1 may mostly 
harbor larger particles. On the other hand the belts around HD\,120326 and HD\,146897 are significantly 
larger than expected using a blackbody assumption suggesting a substantial overabundance of hot small
grains. Interestingly, HD\,146897 is one of those rare F-type stars where CO emission was detected 
\citep{lieman-sifry2016} raising the possibility that the presence of gas may play a role in accumulation 
of small grains in these systems e.g. by retaining those particles that would be blown out by the radiation pressure in 
a gas-free environment \citep[see e.g.][]{wyatt2018}. 
As already noted by \citet{Currie2015}, HD\,115600 exhibits a moderately 
high $\Gamma$ value of $\sim$3.3 indicating the presence of copious amount of small dust particles in the disk.

\begin{figure*}
\begin{center}
\includegraphics[width=0.65\textwidth]{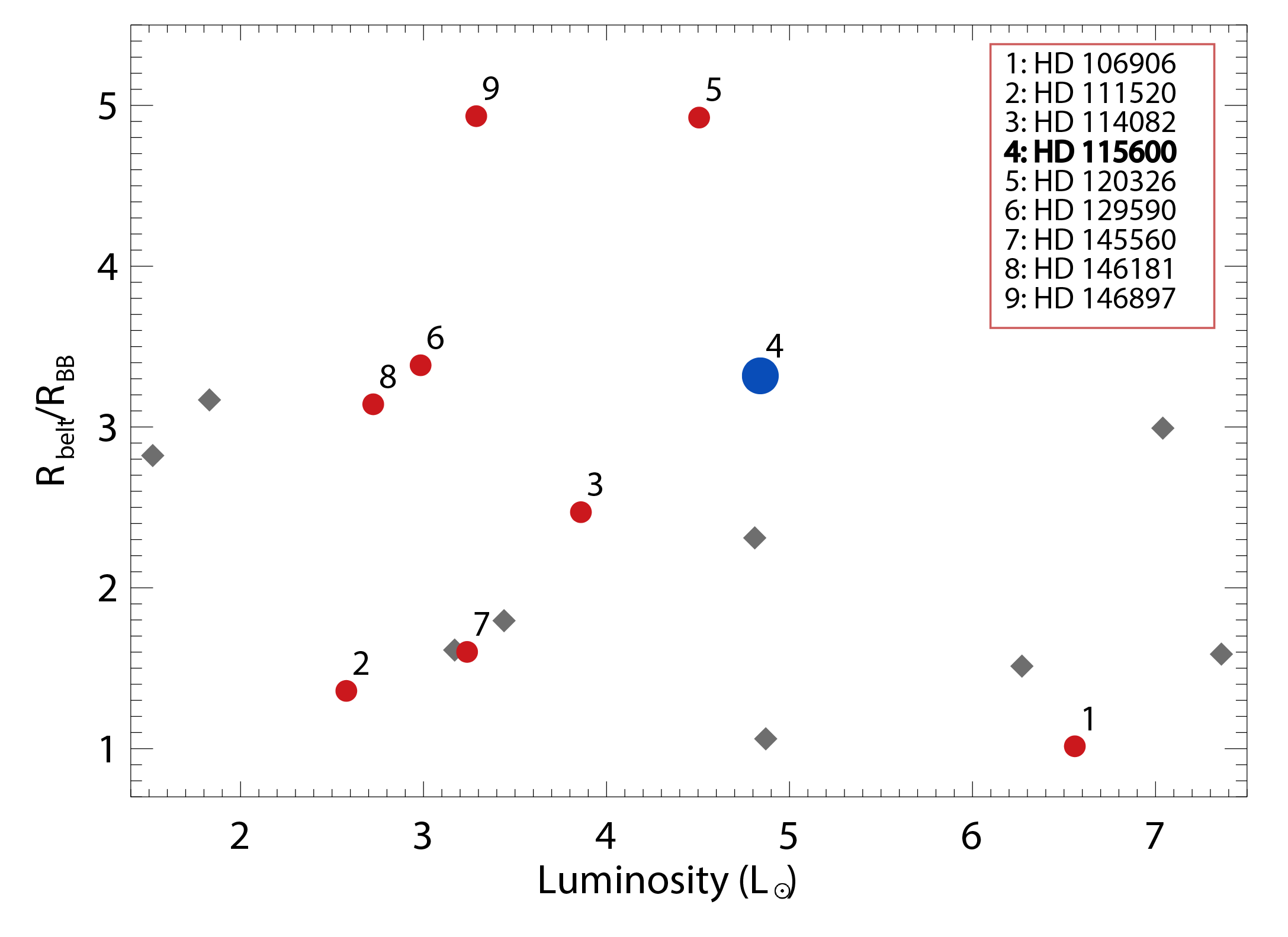}
\end{center}
\caption{The ratio of the measured disk radius to its blackbody radius as a function 
of stellar luminosity. Dust-rich debris disks around Sco-Cen members are displayed by 
red circles while disks from \citet{pawellek2014} are marked by gray diamonds.
\label{fig:gamma}}
\end{figure*}

\section{Spectral Analysis}
\label{sec:spectra} 

We extracted YJ-band spectra from both the processed data, and from our wavelength-dependent disk model that was previously described in Section \ref{sec:analysis}. Both spectra were extracted with a similar methodology to \citet{Currie2015}, to better compare our results. We first measured the average surface brightness, that is the mean pixel value, in each wavelength channel within two 8-pixel diameter apertures centered on the disk extremities in both the IFS and disk model images. These were the same regions selected in Section \ref{sec:results} and by \citep{Currie2015} for spectra, and contain the highest signal-to-noise. We converted these raw surface brightnesses into reflectance spectra (i.e., the relative brightness of the disk compared to HD 115600) by dividing the peak value of the flux calibration PSF in each wavelength, as determined by a Gaussian fit (IDL routine \texttt{GAUSS2DFIT}). The spectra from both disk ansa were then averaged together to produce a combined spectrum. Unfortunately, we found that the spectrum produced from the processed IFS data suffers from significant variability due to speckle noise, and therefore only kept the combined reflectance spectrum from our model for analysis. The uncertainty in the model reflectance spectrum is determined as a combination of the measured speckle noise in each wavelength channel (using the same method from Section \ref{sec:results}), and the brightness uncertainty in the model from the grid step size, discussed in Section \ref{sec:analysis}.

The spectrum of a debris disk depends, among other factors, on the dust composition. By combining our SPHERE YJ-band reflectance spectrum with the GPI H-band reflectance spectrum from \citet{Currie2015}, and comparing them to reflectance models, we aim to constrain the disk composition further. To avoid biases inherent in combining the data from multiple instruments into a single spectrum, we compare the SPHERE spectrum to reflectance models individually. We choose to compare the SPHERE spectrum to the same reflectance models used in \citet{Currie2015}, for water ice, amorphous carbon, and amorphous silicate dust compositions. These models are based on simple Mie scattering theory predictions from \citet{Lisse1998} and Halley-like particle size distributions \citep{Swamy1988,Mazets1986}, rather than more realistic calculations. To convert our reflectance spectrum into normalized units, we scale the spectrum so that the spectrum's maximum is equal to a reflectivity of one. We then scale each reflectance model to minimize $\chi^2$, and the lowest is the best-fit. 

Our reflectance spectrum for SPHERE is shown in Figure \ref{fig:spec}, along with the aforementioned reflectance models. It shows a reddish slope, and, without considering GPI H band spectra yet, has the best agreement with amorphous carbon and silicate compositions, marginally favoring carbon. Water ice is slightly disfavored due to its blue slope, with a $\chi^2$ values 2 times that of amorphous carbon. The uncertainty on the spectrum, however, means that none of the reflectance models are ruled out at the 95\% confidence limit, and any of the 3 compositions could be viable.

Our spectral results in the Y-J bands suggest a more complex composition than the range of those considered in \citet{Currie2015} using the H band. The H band spectrum from GPI data marginally favors water ice composition while disfavoring amorphous carbon, and our Y-J band spectrum from SPHERE data suggests the opposite, however, both spectra could be viable with all our reflectance models. This illustrates the difficulty of obtaining precise composition determinations of debris disks, as the spectra are mostly flat and different compositions are highly degenerate. Of course, the true disk composition is likely to be a complex mix of outer solar system materials (i.e. ices, rock, and complex organics), such that a superposition of spectra may yield a better fit. Comparing the overall Y-H colors of the HD 115600 disk with those of solar system objects can provide some clues to a mixed composition (\citealt{Lisse2017}, Fig. 4). As solar system objects become more active, they tend to become redder in the NIR due to increasing exposure of organic material and loss of surface ices. The active Centaur 5145 Pholus, for example, has an extremely red reflectance spectrum up to about 1.5 microns, but flattens out and becomes slightly blue at longer wavelengths, similar to YJ to H bands the trends we see for the HD 115600 disk. The trend for 5145 Pholus is known to be caused by exposed reddish complex organic materials mixed with surface ice absorption \citep{Barucci2003}. Other active Centaurs display similar behavior. By contrast, highly processed short period comets maintain a reddish color out to 4-5 microns due to their extreme activity and solar processing \citep{Quirico2016}. Interestingly enough,  the approximately 100K dust in the bright ring surrounding A0.5V HR4796A at 75 AU distance matches this behavior \citep{Lisse2017}. On the other hand, pristine Kuiper Belt Objects (KBOs) at T < 50K evince mostly neutral or blue spectra due to the dominance of abundant ices (CO, CH4, N2, H2O) on their surfaces. We can therefore hypothesize that the HD115600 disk has a mixed composition and is composed of  dust emitted from large semi-active planetesimals.

\begin{figure*}
\centering
\includegraphics[width=0.8\textwidth]{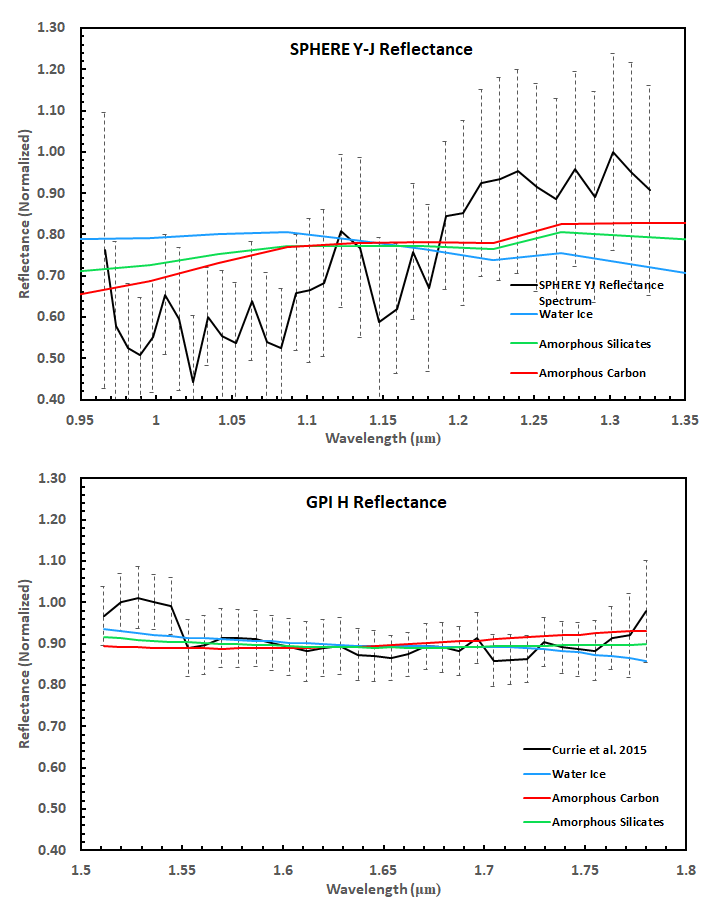}
\caption{\textbf{HD 115600 reflectance spectra.} Reflectance spectra (black) from SPHERE YJ-band (top) and GPI H-band (bottom) compared to Mie theory predictions for water ice, amorphous carbon, and amorphous silicates \citep{Currie2015}. Each reflectance model is shown scaled to best-fit the spectra.}
\label{fig:spec}
\end{figure*}

\begin{figure*}[t]
\centering
\includegraphics[width=0.85\textwidth]{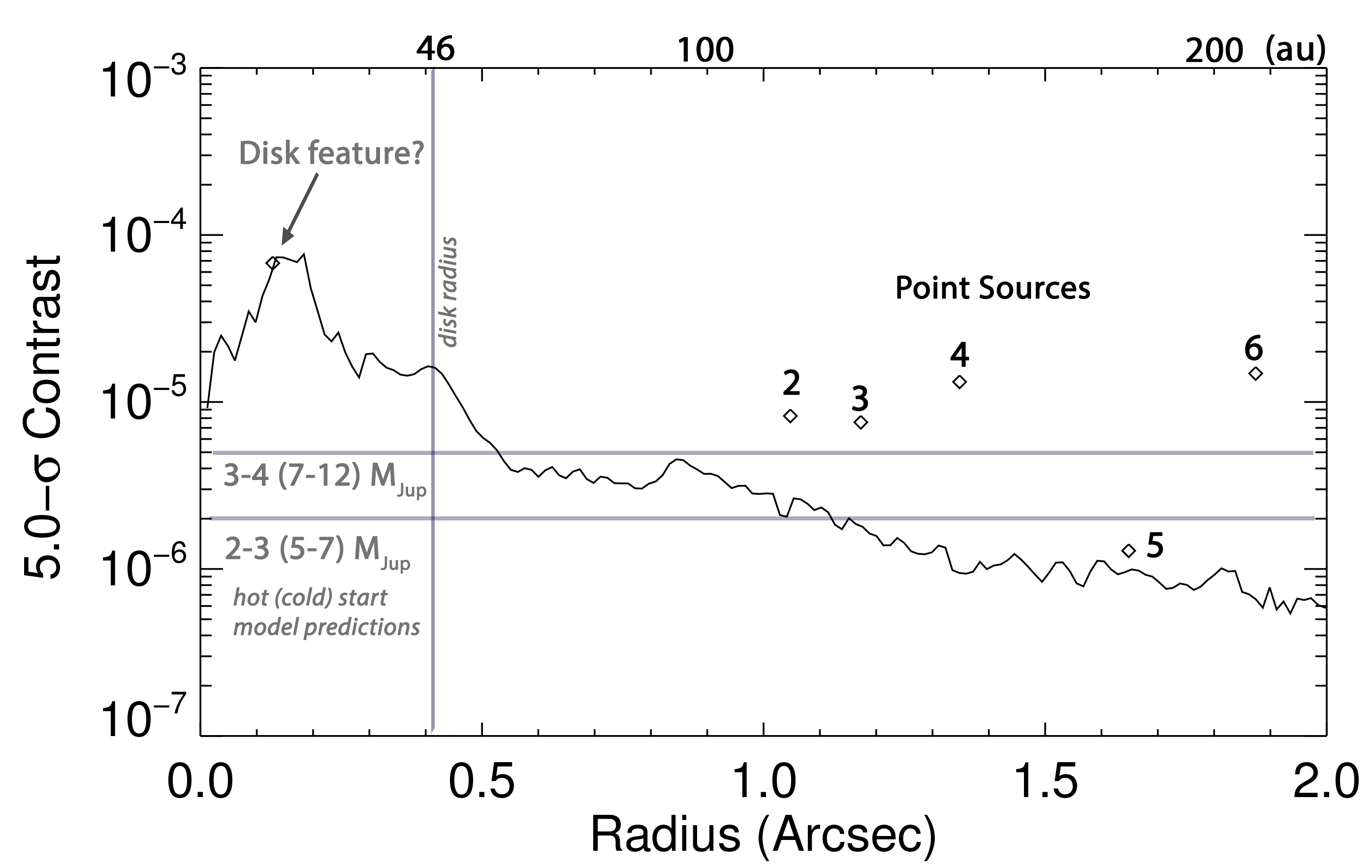}
\caption{\textbf{Radial sensitivity for the KLIP reduction of IRDIS H23 Data.} The disk is clearly present and limits sensitivity around 0.2 arcseconds. Mass sensitivity limits from hot and cold start model predictions \citep{Baraffe2003, Marley2007,Mordasini2013} are shown as horizontal lines. Within the IFS field-of-view($\sim 1$ arcsecond), we achieve a sensitivity of $\sim 3-4 (7-12) M_{Jup}$ for hot (cold) start models. Limits may be slightly higher for particularly dusty, cloudy planets that are slightly underluminous in $H$ band \citep[e.g.][]{Currie2011,Currie2018,DeRosa2016}.  The contrast of several point-sources within the IRDIS field-of-view are numbered.}
\label{fig:sensitivity}
\end{figure*}

\section{Planet Detection Limits}
\label{sec:planets}

While no wide-orbit giant planet is obviously detected in our data, the lack of detection places useful constraints on the presence of planets external to the disk (projected radii $>50$ au). To analyze our detection threshold we injected false-positive point sources at 4$\times 10^{-5}$ contrast into the raw data prior to processing. We then measured the brightness of the recovered sources compared to the originally injected PSF, and utilized the average recovered contrast to estimate the throughput correction appropriate for the data processing. We find that the throughput correction is essentially uniform with radius exterior to the disk, and thus we adopt a constant correction throughout the image. To compute a detection threshold for each radius, we integrated the flux in non-overlapping apertures of radius equivalent to the beam FWHM while accounting for small sample statistics \citep{Mawet2014}. We multiplied the standard deviation of these measurements at a particular radius by the throughput correction factor and scaled this by a factor of five to arrive at the reported $5 \sigma$ contrast limit shown in Figure \ref{fig:sensitivity}. This analysis suggests that we would have confidently detected a point source of 5$\times 10^{-6}$ (H<20.6) between the outer edge of the disk and 100 au, or a source of 2$\times 10^{-6}$ (H<21.6) between 100-200 au. Compared to hot(cold)-start evolutionary models \citep{Baraffe2003, Marley2007,Mordasini2013} which give theoretical planet luminosities at the system's age, these brightness measurements correspond to mass limits of 3-4 (7-12) and 2-3 (5-7) $M_{\rm J}$, respectively, depending on the core mass. 

\section{Summary}
\label{sec:sum}

We present new extreme adaptive optics (VLT/SPHERE) imaging of the debris disk surrounding the young star HD~115600. Our dataset improves characterization of the surrounding debris disk and upper limits on possible exoplanets embedded in the disk. We model the debris disk to investigate structure that could indicate unseen exoplanet(s), as suggested by \cite{Currie2015} and \cite{Thilliez2017}. We determine disk geometry by two methods, using the code ZODIPIC \citep{Kuchner2012} to inject and subtract a grid of disk models into the raw data, and by a 'maximum merit' ellipse fitting method on the ADI images, as described in \cite{Thalmann2011}.  The key findings of our study are as follows:

1) We confidently detect the disk (SNR$\sim 20$) at separations between 0.35\arcsec{} and 0.48\arcsec{} and at wavelength between 1.0 $\mu$m and 1.6~$\mu$m and present near diffraction-limited images at these wavelengths.


2)Using ZODIPIC, we determine a best-fit disk model with central radius $a_0 = 46 \pm 2 $au, inclination $i=80^{\circ} \pm 1$, position angle of $27^{\circ} \pm 1$, and offsets $\Delta \alpha, \Delta \delta$ = $-1.0 \pm 0.5, 0.5 \pm 0.5$ au. By ellipse fitting, we determine a best-fit disk model with projected semimajor axis $a = 44.5 \pm 0.8$ au, inclination $i=78.5^{\circ} \pm 1.0$, position angle of $27.5^{\circ} \pm 1.1$, and offsets of $\Delta \alpha, \Delta \delta$ = $0.8 \pm 0.6, 0.0 \pm 0.6$ au.   

The disk offset is smaller than previous estimates derived from modeling with the H-G scattering function.   As a result, planets responsible for sculpting the debris disk could be significantly lower in mass than previously estimated.

3) Using SPHERE IFS, we produce a YJ-band reflectance spectrum for the disk, which is reddish, and marginally supports an amorphous carbon composition, although other compositions cannot be decisively ruled out. Considering the flat to slightly blue H band reflectance previously seen with GPI and a survey of our solar system's small body reflectance spectra, the combined results are consistent with a mixed organic and water ice dust composition typical of our solar system's active Centaurs interacting with the giant planets and the Sun.

4) We constrain a mass sensitivity limit of 3-4(7-12) $M_{\rm J}$ with 100 au of the star, and 2-3(5-7) $M_{\rm J}$ within 100-200 au by combining contrast limits from fake planet injections with hot(cold)-start model predictions.


\appendix
\section{Properties of detected point-sources}

To investigate the nature of the sources in the SPHERE and GPI images, we list the astrometry in tables 4 and 5, and the astrometric difference of commonly detected sources in table 6. Note that source 1 is a likely disk feature as it falls within the region of the scattered light ring. Sources 2-4 exhibit motions in a consistent direction with one another, and consistent with the reflex motion of the proper motion of HD 115600, and thus we classify these sources as background stars. 

\begin{deluxetable}{ccccc}[H]

\tablecaption{Properties of Point-Sources detected around HD 115600 in the IRDIS Field}


\tablehead{\colhead{Source}  & \colhead{SNR} & \colhead{Contrast} & \colhead{$\rho$ (")} & \colhead{$\theta$ (deg)} }

\startdata
       1    &   5.81 & 6.78$\times 10^{-5}$  &   0.1284  &    53.69 \\
       2      &  14.69 & 8.25$\times 10^{-6}$  &    1.0476  &    211.60\\
       3      &  22.14 & 7.56$\times 10^{-6}$  &    1.1727  &    63.03\\
       4      &  47.91 & 1.32$\times 10^{-5}$  &    1.3481  &    202.95\\
       5    &  6.81 & 1.29$\times 10^{-6}$  &    1.6482  &    144.53\\
       6     &  76.50 & 1.48$\times 10^{-5}$  &    1.8739  &    222.25\\
       7     &  23.77 & 2.67$\times 10^{-6}$  &    2.1634  &    272.49\\
       8     &  5.11 & 6.04$\times 10^{-7}$  &    2.4740  &    259.48\\
       9      &  9.51 & 1.04$\times 10^{-6}$  &    2.6082  &    290.10\\
      10      &  11.66 & 1.26$\times 10^{-6}$  &    2.6159  &    103.44\\
      11    &  53.06 & 6.48$\times 10^{-6}$  &    2.8354  &    306.45\\
      12      &  420.11 & 5.05$\times 10^{-5}$  &    2.8929  &    3.56\\
      13     &  290.13 & 3.51$\times 10^{-5}$  &    2.9851  &    217.98\\
      14      &  36.92 & 3.30$\times 10^{-6}$  &    3.2857  &    69.59\\
      15      &  9.65 & 8.88$\times 10^{-7}$  &    3.3603  &    322.31\\
      16     &  17199.3 &  0.00154  &    3.3732  &    63.39\\
      17     &  17.74 & 2.14$\times 10^{-6}$  &    3.4622  &    209.25\\
      18      &  110.93 & 2.65$\times 10^{-5}$  &    3.5305  &    339.53\\
      19     &  221.18 & 4.08$\times 10^{-5}$  &    3.5837  &    190.02\\
      20      &  7.70 & 1.13$\times 10^{-6}$  &    3.6186  &    328.37\\
      21      &  22.01 & 2.25$\times 10^{-6}$  &    3.7714  &    104.46\\
      22     &  7.54 & 7.95$\times 10^{-7}$  &    3.7861&      245.12\\
      23      &  94.40 & 9.20$\times 10^{-6}$  &    3.9049&      210.63\\
      24      &  6.02 & 7.07$\times 10^{-7}$  &    3.9917 &     284.77\\
      25     &  11.50 & 1.32$\times 10^{-6}$  &    3.9924  &    301.10\\
      26      &  111.28 & 1.47$\times 10^{-5}$  &    4.0119  &    134.29\\
      27      &  68.16 & 6.44$\times 10^{-6}$  &    4.0512  &    122.57\\
      28     &  6.21 & 6.32$\times 10^{-7}$  &    4.1346  &    208.78\\
      29     &  89.68 & 8.20$\times 10^{-6}$  &    4.2913  &    155.16\\
      30     &  56.24 & 5.10$\times 10^{-6}$  &    4.6783  &    316.74\\
      31      &  190.46 & 2.12$\times 10^{-5}$  &    4.7166  &    38.88\\
      32    &  130.87 & 1.18$\times 10^{-5}$  &    4.9489  &    45.61\\
      33     &  3203.20 & 0.0002726  &    5.0191  &    88.65\\
      34     &  7.90 & 6.53$\times 10^{-7}$  &    5.0434  &    353.76\\
\enddata


\tablecomments{Column(1): Sources as numbered in Figure \ref{fig:irdis} and Figure \ref{fig:sensitivity}. Column(2): Detection significance. Column(3): Contrast to HD 115600. Column(4): Separation in arcseconds. Column(5): Position angle in degrees.  }
\label{tab:props}
\end{deluxetable}

\begin{deluxetable}{cccc}[H]

\tablecaption{Properties of Point-Sources detected around HD 115600 in the GPI Field}


\tablehead{\colhead{Source}  & \colhead{SNR}  & \colhead{$\rho$ (")} & \colhead{$\theta$ (deg)} }

\startdata 
	   2    &   8.070  &   1.0854  &    213.18 \\
       3    &   8.15  &   1.1488   &    62.94 \\
       4    &   8.99  &   1.3737   &    204.12 \\
\enddata


\label{tab:props}
\end{deluxetable}

\begin{deluxetable}{ccc}[H]

\tablecaption{Astrometric Motion of Commonly Detected Sources (Relative to HD 115600 over 2015.42-2014.31)}


\tablehead{\colhead{Source}  & \colhead{$\Delta$ RA (mas)} & \colhead{$\Delta$ Dec (mas)} }

\startdata
       2     &   45.1 &    16.2 \\
       3      &   22.1   &    9.3 \\
       4      &   35.6  &    12.4 \\
       Mean & 34.2 & 12.6 \\
       StdDev & 11.5 & 3.4 \\
       \hline
       HD 115600 & PM RA & PM Dec\\
       GAIA DR2 & -32.63$\pm$0.05 mas/yr & -18.32$\pm$0.05 mas/yr\\
\enddata



\label{tab:props}
\end{deluxetable}

\acknowledgments

\software{IDL}

 \newcommand{\noop}[1]{}

\end{document}